\documentclass[epj]{webofc}
\usepackage[varg]{txfonts}   
\wocname{EPJ Web of Conferences}
\woctitle{CONF12}
\begin{document}
\selectlanguage{english}
\title{Recent results from the NA61/SHINE strong interaction physics programme}
%
%

\author{Evgeny Andronov\inst{1}\fnsep\thanks{\email{e.v.andronov@spbu.ru}}, for the NA61/SHINE Collaboration 
}

\institute{Saint Petersburg State University, ul. Ulyanovskaya 1, 198504, Petrodvorets, Saint Petersburg, Russia
}

\abstract{%
  The main physics goals of the NA61/SHINE programme on strong interactions are the study of the properties of the onset of deconfinement and the search for signatures of the critical point of strongly interacting matter. For this purpose a scan of the two dimensional phase diagram ($T$-$\mu_{B}$) is being performed at the SPS by measurements of hadron production in nucleus-nucleus collisions as a function of collision energy and system size.

This contribution presents intriguing results on the energy dependence of hadron spectra and yields in inelastic p+p and centrality selected Be+Be and Ar+Sc collisions. In particular, the energy dependence of the signals of deconfinement, the "horn", "step" and "kink", and new results on fluctuations and correlations are shown and compared with the corresponding data of other experiments and model predictions.
}
\maketitle
\section{Introduction}
\label{intro}
\begin{figure}[h]
\centering
\includegraphics[scale=.25]{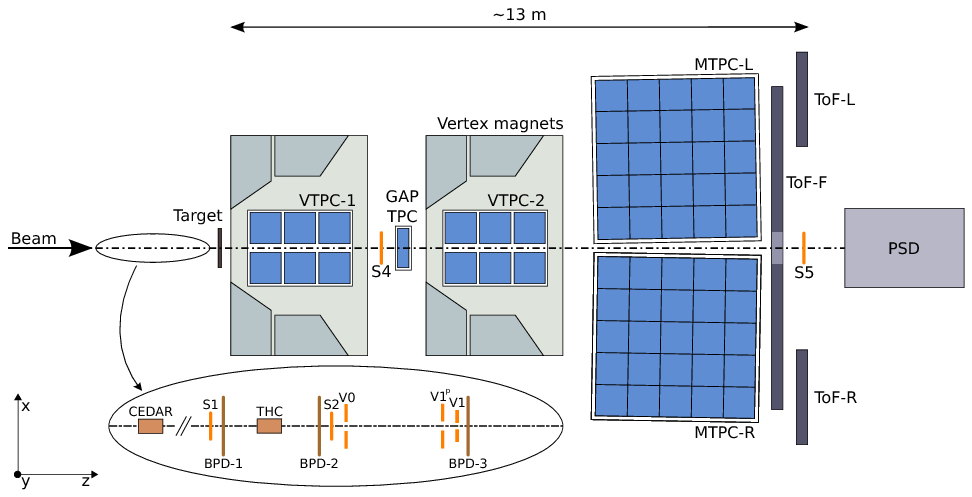}
\caption{Schematic layout of the NA61/SHINE experiment at the CERN SPS
(horizontal cut in the beam plane, not to scale).}
\label{fig1} 
\end{figure}

NA61/SHINE~\cite{NA61} is a fixed target experiment at the Super Proton Synchrotron (SPS) of the European Organization for Nuclear Research (CERN). The layout of the NA61/SHINE detector is sketched in Fig.~\ref{fig1}.
It consists of a large acceptance hadron spectrometer with
excellent capabilities in charged particle momentum measurements and
identification by a set of five Time Projection Chambers as well as
Time-of-Flight detectors.
The geometrical layout of the TPCs allows particle detection down to $p_{T}=0$ GeV/c in a broad  interval of the forward rapidity semisphere, which is practically impossible at collider experiments. 
The high resolution, modular forward calorimeter,
the Projectile Spectator Detector, measures forward going energy $E_{F}$, which in nucleus-nucleus reactions is primarily a measure of
the number of spectator (non-interacted) nucleons and thus
related to the centrality of the collision. 

\begin{figure}[h]
\centering
\includegraphics[width=5cm,clip]{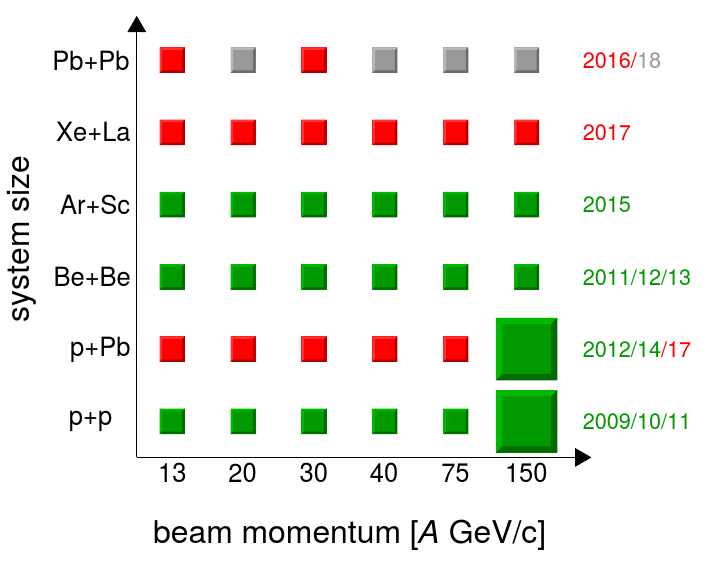}
\caption{Data taking status of the strong interaction programme of NA61/SHINE.}
\label{datatak}       
\end{figure}

The main goal of the strong interaction programme of the experiment is to discover the Critical Point (CP)~\cite{Fodor:2004nz}  of strongly interacting matter
and study the properties of the onset of deconfinement (OD)~\cite{Gazdzicki:1998vd, Alt:2007aa}. To achieve this goal a two-dimensional phase diagram scan - energy versus system size - is being performed by NA61/SHINE. Both primary and secondary beams are available to the experiment, allowing measurements of hadron production in collisions of protons and various nuclei (p+p, Be+Be, Ar+Sc, Xe+La) at a range of beam momenta (13{\it A} - 158{\it A} GeV/c). Figure~\ref{datatak} shows for which systems and energies data has already been collected (green), is
scheduled for recording (red) or is planned (gray). This scan allows to probe different values of temperature $T$ and baryochemical potential $\mu_{B}$ of the strongly interacting matter at the freeze-out stage~\cite{Becattini:2006}.
\section{Spectra and yields: studying the properties of the onset of deconfinement}
\label{spec}
\begin{figure}[h]
\centering
\includegraphics[width=4.5cm, trim={6cm 11cm 6cm 11cm},clip]{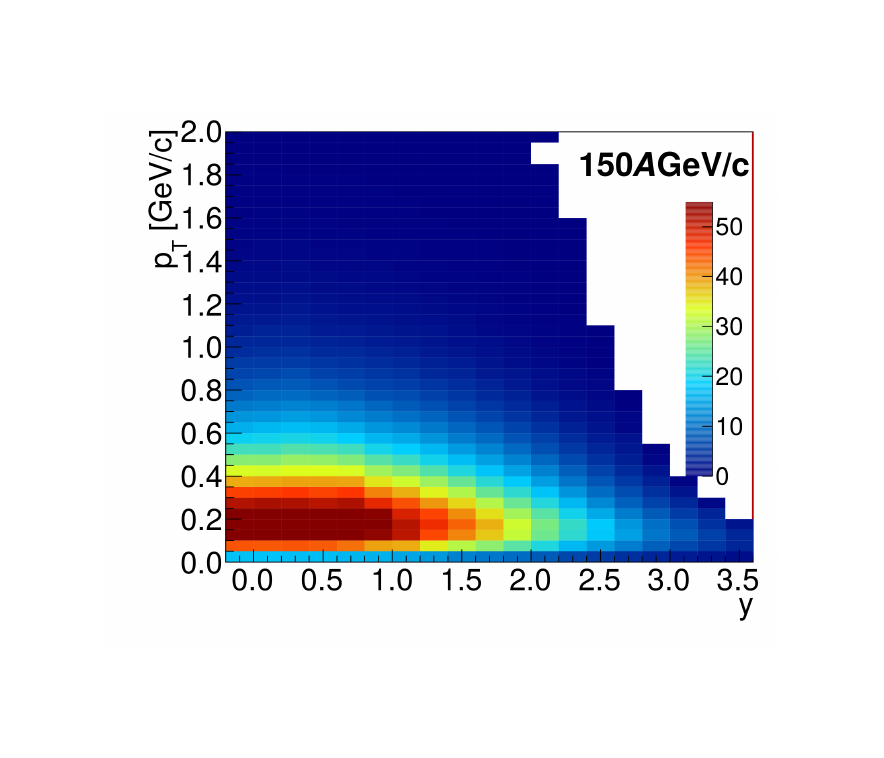}
\includegraphics[width=4.5cm, trim={6cm 11cm 6cm 11cm},clip]{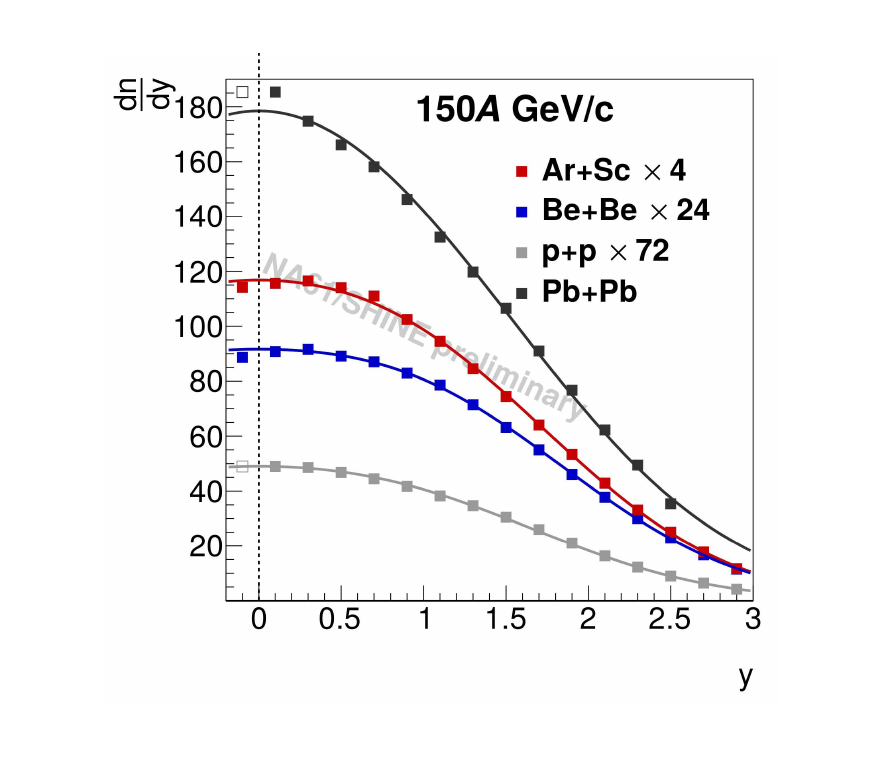}
\includegraphics[width=4.5cm, trim={6cm 11cm 6cm 11cm},clip]{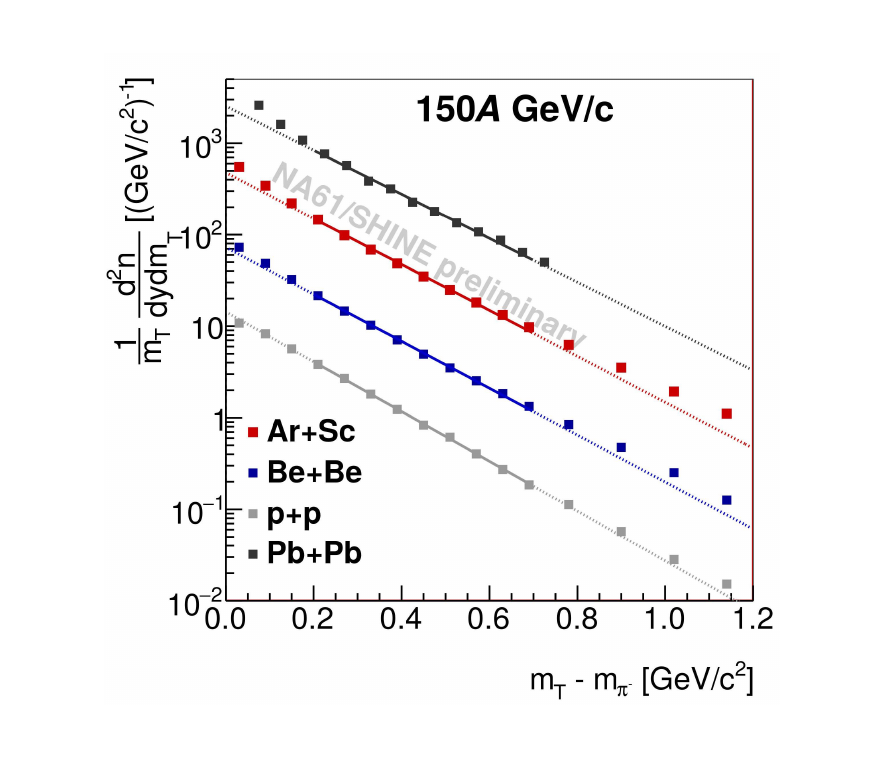}
\caption{Preliminary $\pi^{-}$ spectra for 5$\%$ of events with the lowest $E_{F}$ in Ar+Sc collisions at 150{\it A} GeV/c. Only statistical uncertainties are shown. Left: double differential spectra $\dfrac{dn^{2}}{dydp_{T}}$. Middle: spectra of rapidity $y$ for central Ar+Sc collisions compared to NA49 results from central Pb+Pb collisions and NA61/SHINE results from inelastic  p+p and central Be+Be collisions. Right: spectra of transverse mass $m_{T}$ for the same systems, with lines representing exponential fits in $\left(m_{T}-m_{{\pi}^{-}}\right)\in\left(0.2;0.7\right)$}
\label{spectra}       
\end{figure}

One of the main signals of the onset of deconfinement are the kink, horn, and step~\cite{Gazdzicki:1998vd} structures observed in Pb+Pb collisions by the NA49 collaboration. 
Analysis of spectra and yields by NA61/SHINE allows to check whether these structures are present also in collisions of small and intermediate mass nuclei. Recent measurements of Argon on Scandium collisions are an important step in this program.
Figure~\ref{spectra} shows the spectra of $\pi^{-}$ from strong and electromagnetic processes in Ar+Sc collisions at 150{\it A} GeV/c obtained with the $h^{-}$ analysis method~\cite{Lewicki:2016}. The fact that approximately 90$\%$ of negatively charged hadrons produced in the SPS energy range are $\pi^{-}$ mesons is at the core of this method. Correction for contribution from other negatively charged particles is done using Monte Carlo simulations based on the EPOS 1.99 model~\cite{Pierog:2009zt} together with the GEANT-3.2 code for particle transport and detector simulation.

\begin{figure}[h]
\centering
\includegraphics[width=5cm,clip]{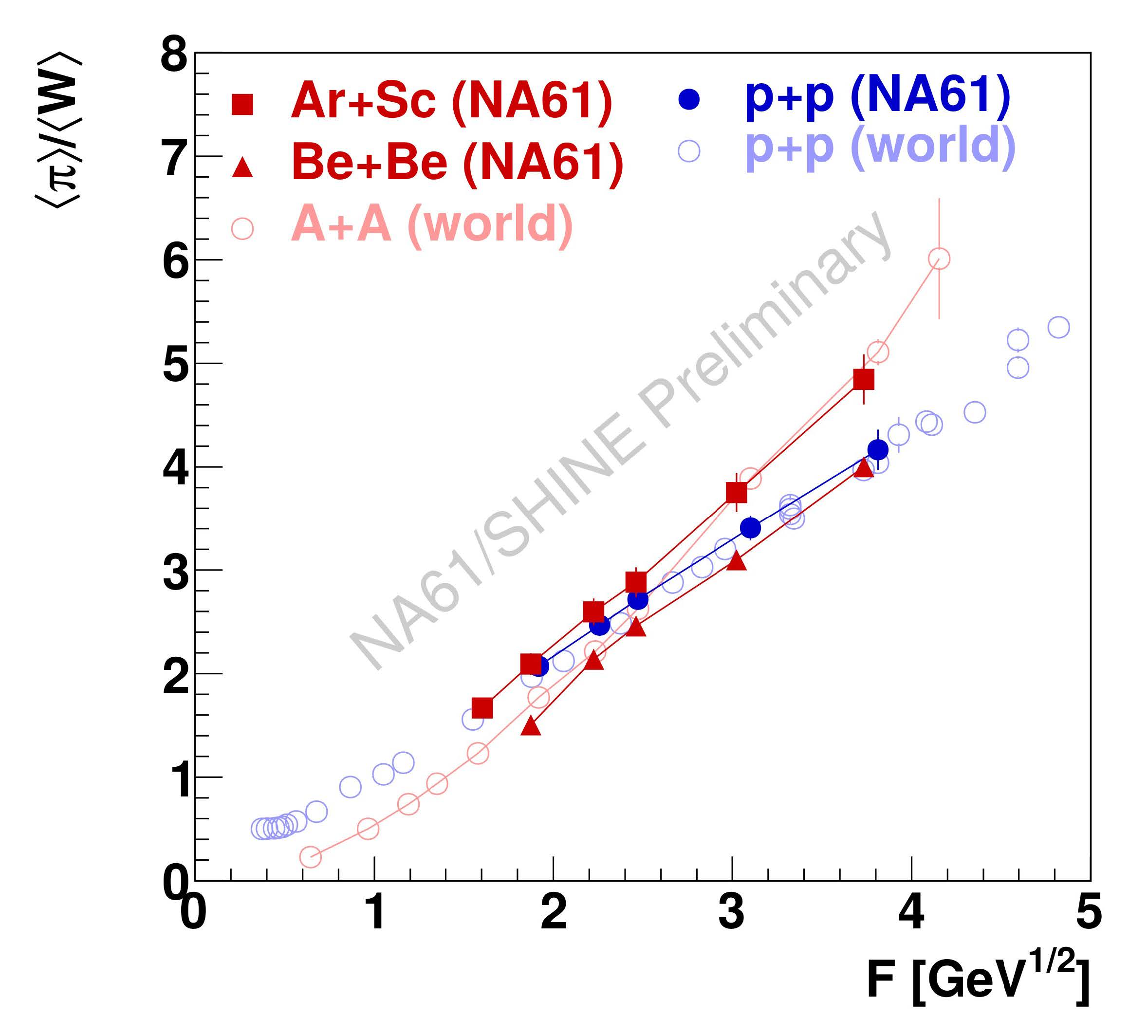}
\caption{Preliminary results for the mean pion multiplicity ${\langle\pi\rangle}$ divided by the mean number of wounded nucleons $\langle W\rangle$ versus the Fermi energy measure $F\simeq s^{1/4}_{NN}$ for inelastic p+p interactions and central Be+Be, Ar+Sc collisions from NA61/SHINE and for world data on p+p and central A+A collisions~\cite{Golokhvastov:2001,Abbas:2013}.}
\label{kink}       
\end{figure}

Rapidity spectra (see Fig.~\ref{spectra}, middle) were fitted to obtain $4\pi$ mean multiplicities of $\pi^{-}$ mesons~\cite{Naskret:2016}. As the measurements via the $h^{-}$ analysis method are possible only for $\pi^{-}$ mesons, the multiplicities of $\pi^{+}$ and $\pi^{0}$ mesons were approximated by ${\langle\pi\rangle}_{p+p}=3{\langle\pi^{-}\rangle}_{p+p}+1$ and ${\langle\pi\rangle}_{A+A}=3{\langle\pi^{-}\rangle}_{A+A}$. In order to compare matter created in the collisions of different nuclei the mean pion multiplicity $\langle \pi\rangle$ is divided by the mean number of wounded nucleons $\langle W\rangle$ corresponding to the given class of collision centrality. This quantity was obtained using the Monte Carlo model - Glissando 2.73~\cite{Broniowski:2007nz}. Figure~\ref{kink} shows the kink plot, where the $\langle \pi\rangle$ multiplicity, normalized to $\langle W\rangle$, increases faster with $F={\left(\frac{{\left(\sqrt{s_{NN}}-2m_{N}\right)}^{3}}{\sqrt{s_{NN}}}\right)}^{1/4}$ in the SPS energy range for central Pb+Pb collisions than in p+p interactions. This behaviour violates the prediction of the Wounded Nucleon Model~\cite{Bialas:1976ed} $\langle\pi\rangle_{A+A}/\langle W\rangle = \langle\pi\rangle_{p+p}/2$, but is successfully explained by the entropy increase due to formation of quark-gluon plasma in the Statistical Model of the Early Stage (SMES)~\cite{Gazdzicki:2010iv}. The new results obtained for central Ar+Sc collisions follow the Pb+Pb trend for high SPS energies and are close to the p+p results for low SPS energies whereas the new results for Be+Be show the opposite tendency. One should mention that the mean number of wounded nucleons $\langle W\rangle$ is a model-dependent quantity, leading to uncertainties when comparing results obtained for different systems.

\begin{figure}[h]
\centering
\includegraphics[width=10cm,clip]{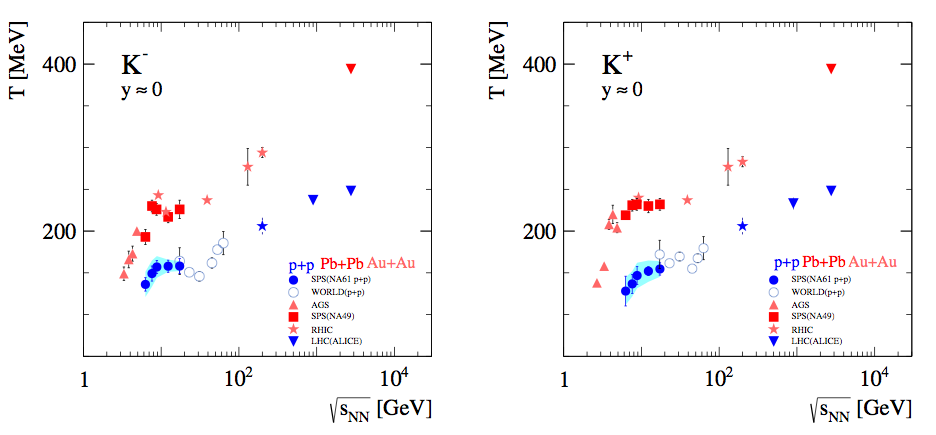}
\caption{Energy dependence of the inverse slope parameter T of transverse mass spectra of K$^-$ and K$^+$ in inelastic p+p interactions measured by the NA61/SHINE experiment (full blue circles) and other experiments (open blue circles) and central Au+Au and Pb+Pb interactions. Blue band represents the systematic uncertainty.}
\label{step}       
\end{figure}

The NA49 collaboration observed a plateau (step) in the inverse slope parameter ($T$) of transverse mass ($m_{T}$) spectra of kaons for Pb+Pb collisions as expected from the SMES model for constant temperature and pressure in a mixed phase. The recent NA61/SHINE results~\cite{Pulawski:2015tka}, presented in Fig.~\ref{step}, show that even in p+p collisions the energy dependence of $T$ for kaons (identified using the $dE/dx$ method) exhibits a rapid change in the SPS energy range.

\begin{figure}[h]
\centering
\includegraphics[width=10cm, trim={0cm 10cm 0cm 10cm},clip]{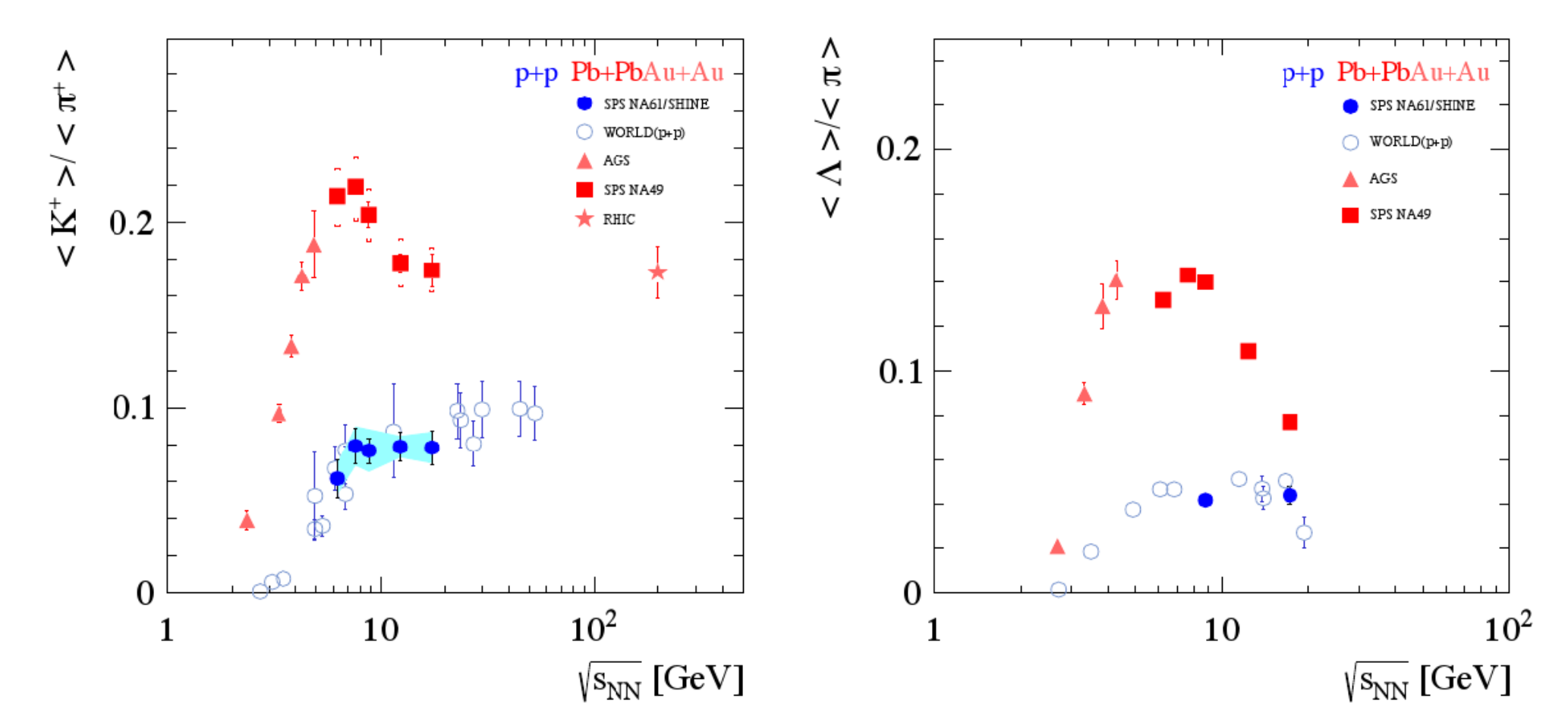}
\caption{Energy dependence of the ratios $\langle K^{+}\rangle/\langle\pi^{+}\rangle$ and $\langle\Lambda\rangle/\langle\pi\rangle$ in inelastic p+p interactions measured by the NA61/SHINE experiment (full blue circles) and world data on p+p and central A+A collisions. Blue band represents the systematic uncertainty.}
\label{horn}       
\end{figure}

Moreover, sharp peaks in the energy dependence of the ratios $K^{+}/\pi^{+}$ and $\Lambda/\pi$ were found for Pb+Pb collisions by the NA49 collaboration. Figure~\ref{horn} shows a comparison of the new measurements by NA61/SHINE for inelastic p+p interactions with the world data. Candidates of charged decays of Lambda hyperons were identified by the standard topological cuts applied to pairs of positively and negatively charged particles detected by the TPCs~\cite{Aduszkiewicz:2015dmr,Stroebele:2016}. One observes that even in p+p interactions the $K^{+}/\pi^{+}$ ratio exhibits rapid changes with energy whereas new measurements of the $\Lambda/\pi$ ratio were done only for two energies and do not give a clear picture of the energy dependence.

\section{Fluctuation observables: search for the critical point}
\label{fluc}
The strategy of looking for the critical point (CP) of strongly interacting matter is based on the expectation that the correlation length $\xi$ diverges at the CP. This divergence may lead to the growth of fluctuations for different observables such as multiplicity, net charge etc. Therefore, one can expect that a scan over freezeout points close to the CP will show non-monotonic behavior of these fluctuation observables.

This search is complicated by the fact that the size of the system created in collisions of two nuclei changes significantly from event to event. Observables can be classified according to their dependence on this volume and its fluctuations: 1) extensive quantity - proportional to the system volume in the Grand Canonical Ensemble or the number of the wounded nucleons in the Wounded Nucleon Model~\cite{Bialas:1976ed} 2) intensive quantity - independent of the system volume 3) strongly intensive quantity~\cite{Gorenstein:2011vq} - independent of the system volume and fluctuations of this volume. Strongly intensive quantities are best suited to study fluctuations in nucleus-nucleus collisions because of the unavoidable event-to-event variations of the volume. 
\subsection{Scaled variance of multiplicity}
\label{scvar}
The most common way to characterize the strength of fluctuations of the quantity $A$ is to determine the variance of its distribution $Var\left(A\right)$ which is an extensive quantity. In turn, the scaled variance
\begin{equation}
\omega[A]=\frac{Var\left(A\right)}{\langle A\rangle}
\end{equation}
is intensive. Here, $\langle \cdots\rangle$ stands for averaging over all events.

In the Wounded Nucleon Model one can get the following expression for the scaled variance of the distribution of the multiplicity $N$:
\begin{equation}
\omega[N]={\omega}^{*}[N]+\langle N\rangle/\langle W\rangle \cdot\omega[W],
\end{equation}
where ${\omega}^{*}[N]$ is the scaled variance calculated for the fixed value $\langle W\rangle$ (i.e. ${\omega}^{*}[N]=\omega[N]_{p+p}$). This leads to the following ineqality
\begin{equation}
\omega[N]_{A+A}\geq \omega[N]_{p+p}
\label{ineq}
\end{equation}

\begin{figure}[h]
\centering
\includegraphics[width=6.5cm,clip]{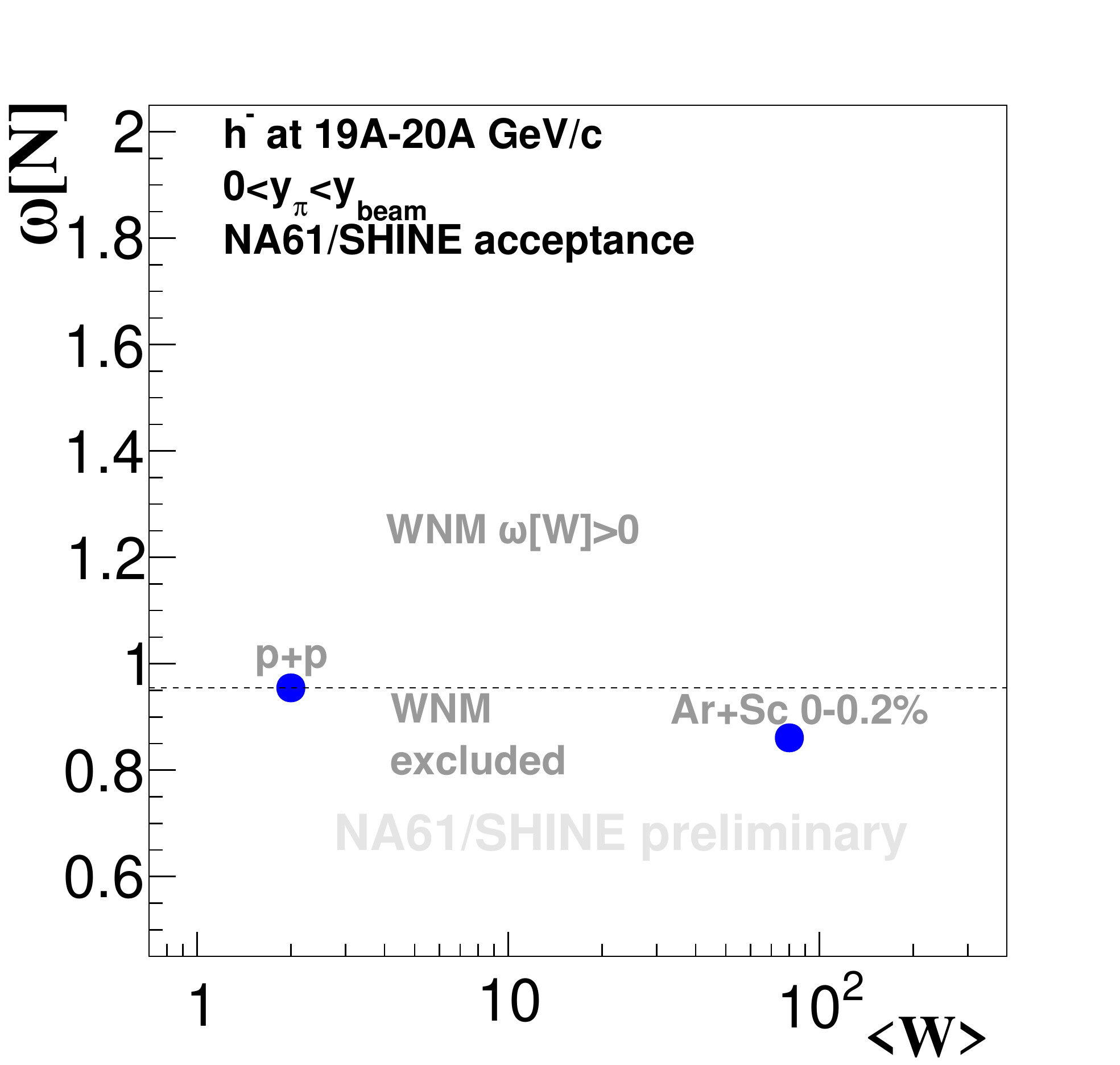}
\includegraphics[width=6.5cm,clip]{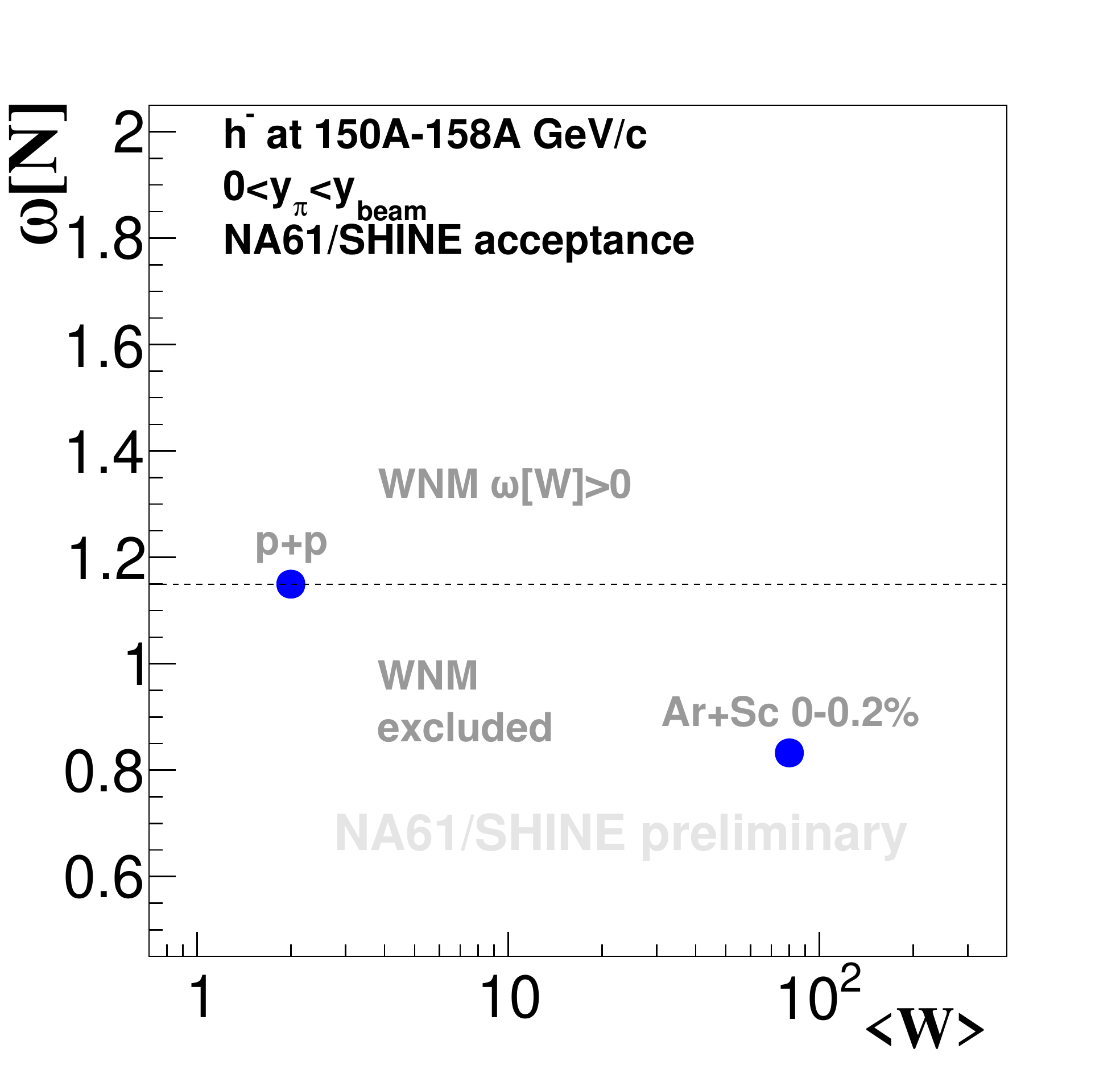}
\caption{Preliminary results on $\omega[N]$ of the multiplicity distribution of negatively charged hadrons in inelastic p+p interactions and for the 0.2$\%$ of events with the lowest $E_{F}$ in Ar+Sc collisions at 19 and 150{\it A} GeV/c. Statistical uncertainties are smaller than the symbol size. The horizontal dashed line divides the plot into WNM exluded (below the line) and WNM allowed (above the line) zones.}
\label{scaled}       
\end{figure}

The recent measurements by the NA61/SHINE collaboration show that this inequality is violated~\cite{Seryakov:2016}. Figure~\ref{scaled} shows the system size dependence of the scaled variance $\omega[N]$ for negatively charged hadrons for two beam momenta - 19 and 150{\it A} GeV/c. One observes that fluctuations of multiplicity in very central Ar+Sc collisions are suppressed compared to p+p interactions, in disagreement with Eq.~(\ref{ineq}).
\subsection{Joint fluctuations of multiplicity and transverse momentum}
\label{joint}
In order to suppress contributions from these 'trivial' fluctuations strongly intensive observables are used: 
\begin{equation}
\Delta[A,B] = \frac{1}{C_{\Delta}} \biggl[ \langle B \rangle \omega[A] -
                        \langle A \rangle \omega[B] \biggr] 
\end{equation}
\begin{equation}
\Sigma[A,B] = \frac{1}{C_{\Sigma}} \biggl[ \langle B \rangle \omega[A] +
                        \langle A \rangle \omega[B] - 2 \bigl( \langle AB \rangle -
                        \langle A \rangle \langle B \rangle \bigr) \biggr]
\end{equation}
In case of joint transverse momentum $P_{T}$ and multiplicity $N$ fluctuations we define:
$A=P_{T}=\sum_{i=1}^{N}p_{T_{i}}$, $B=N$, $C_{\Delta}=C_{\Sigma}=\langle N\rangle \omega[p_{T}]$.

\begin{figure}[h]
\centering
\includegraphics[width=6.5cm,clip]{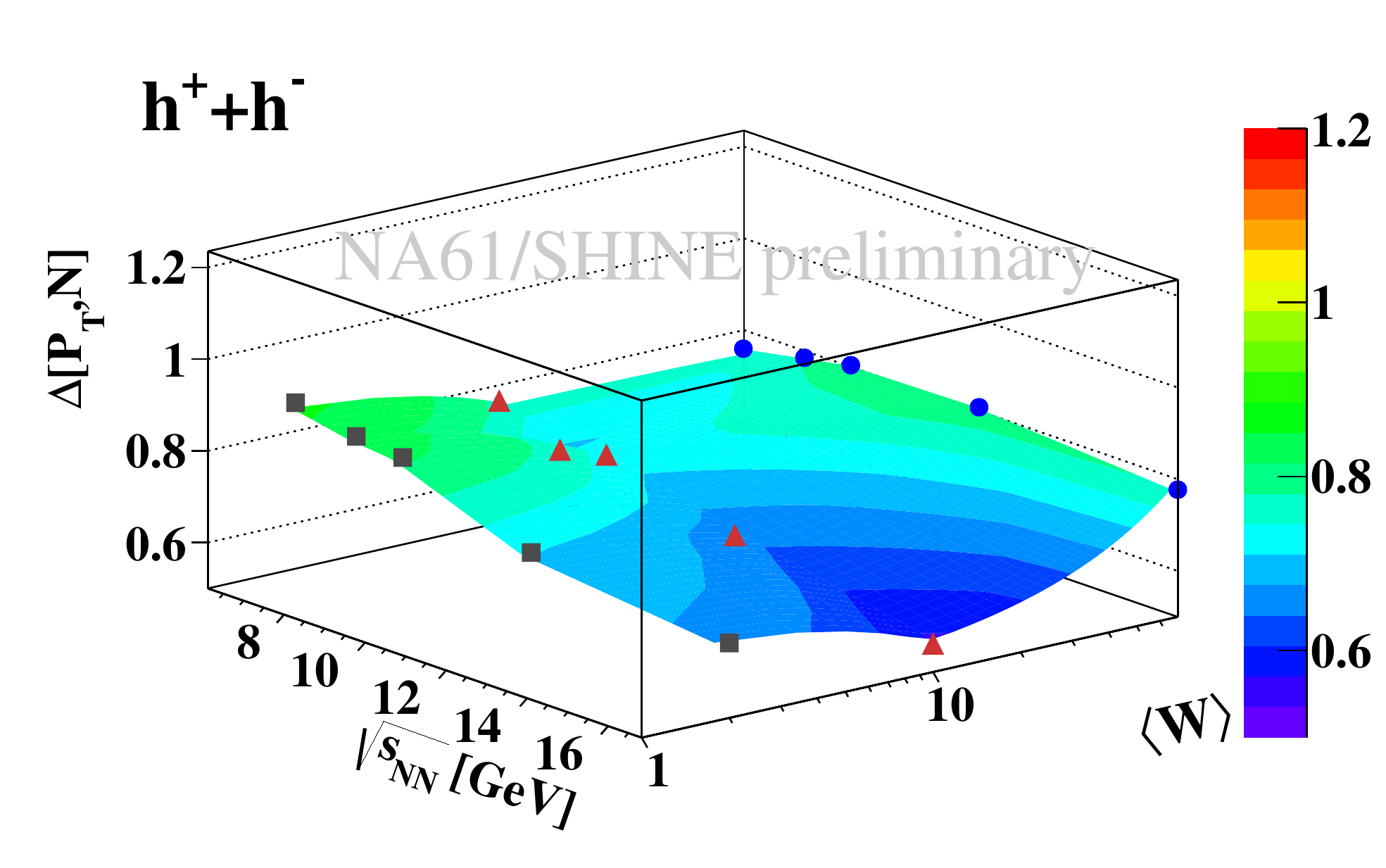}
\includegraphics[width=6.5cm,clip]{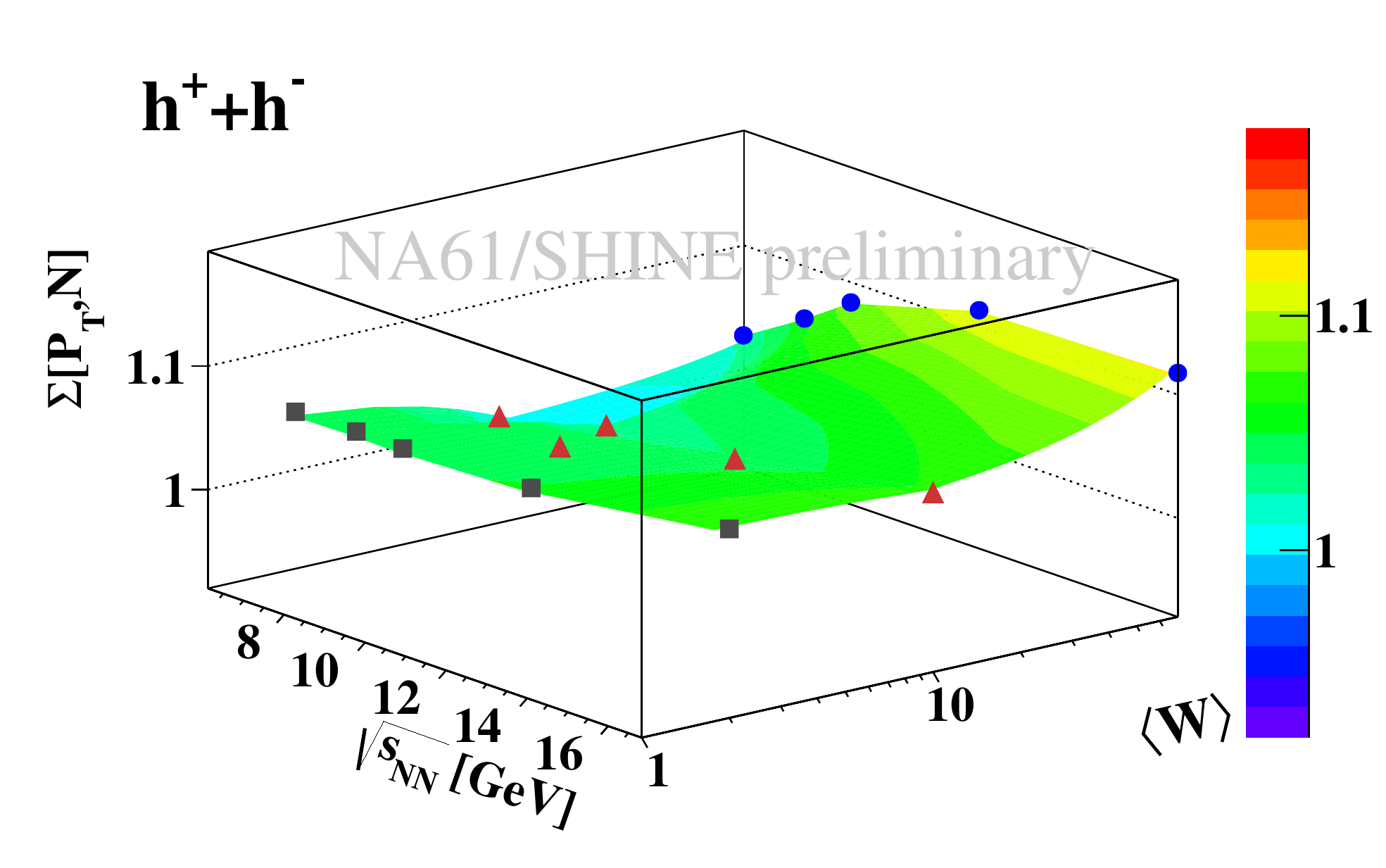}
\caption{Preliminary results on $\Delta[P_{T},N]$ (left) and $\Sigma[P_{T},N]$ (right) of all charged hadrons in inelastic p+p and central Be+Be and Ar+Sc collisions. Statistical uncertainties are smaller than the symbol size.}
\label{deltasigma}       
\end{figure}

The NA61/SHINE collaboration performed measurements of these two quantities for inelastic p+p interactions, as well as Be+Be and Ar+Sc collisions for the smallest 5$\%$ of forward energies~\cite{Andronov:2016}. Figure~\ref{deltasigma} shows $\Delta[P_{T},N]$ (left) and $\Sigma[P_{T},N]$ (right) as a function of $\sqrt{s_{NN}}$ and mean number of wounded nucleons $\langle W\rangle$ (from~\cite{Broniowski:2007nz}). No dip-like or hill-like structures that could be related to the critical point of strongly interacting matter are observed in these results.
\subsection{Higher order moments of the net-charge distribution}
The fluctuation quantities analyzed in sections~\ref{scvar} and~\ref{joint} include only first and second moments of the studied observables. Higher order moments of fluctuations should depend more sensitively on $\xi$, possibly making the signal of the CP more visible. Another reason to measure higher order moments of the net electric charge distribution is that for conserved quantum numbers there is a possibility to compare with susceptibilities calculated in lattice QCD~\cite{Karsch:2011}. The volume independent combinations of the higher order moments are selected as
\begin{equation}
S\sigma = \frac{{\langle Q^{3}\rangle}_{c}}{Var(Q)}, \,\,\,\,\,\,\, \kappa\sigma^{2} = \frac{{\langle Q^{4}\rangle}_{c}}{Var(Q)},
\end{equation}
where ${\langle Q^{3}\rangle}_{c}$ and ${\langle Q^{4}\rangle}_{c}$ are the third and fourth order cumulants of the net charge distribution. 

\begin{figure}[h]
\centering
\includegraphics[width=4.5cm, trim={5cm 11cm 6cm 11cm},clip]{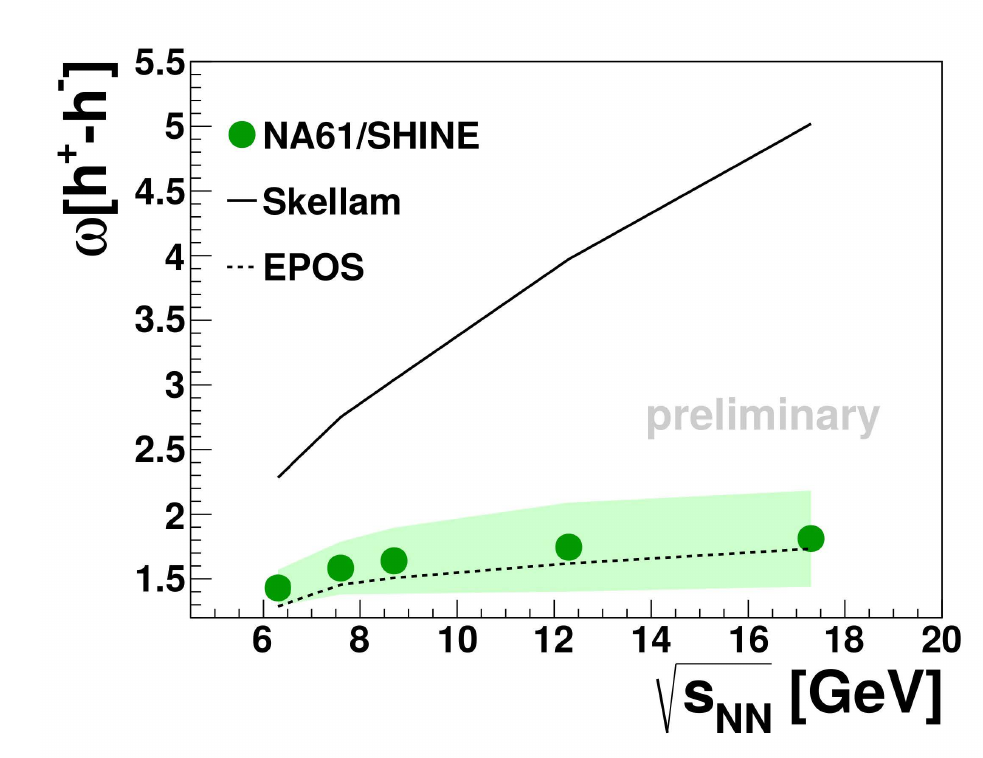}
\includegraphics[width=4.5cm, trim={5cm 11cm 6cm 11cm},clip]{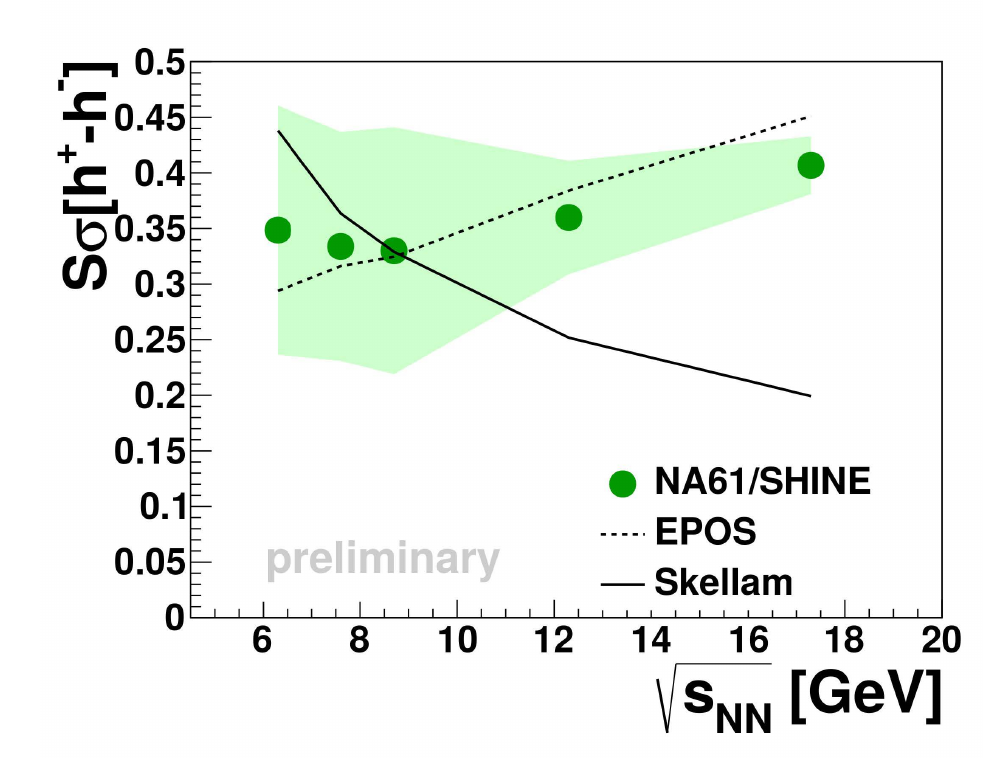}
\includegraphics[width=4.5cm, trim={5cm 11cm 6cm 11cm},clip]{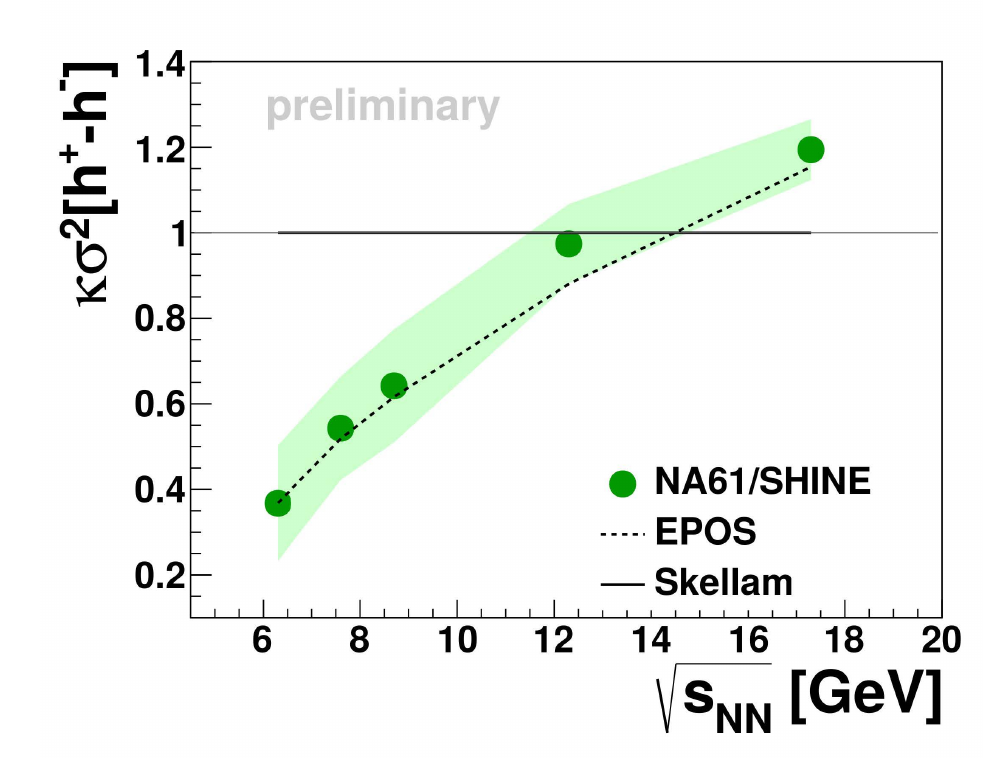}
\caption{Preliminary results on $\omega$, $S\sigma$ and $\kappa\sigma^{2}$ of the net charge distribution in inelastic p+p interactions. Statistical uncertainties are smaller than the symbol size. Green bands represent systematic uncertainties.}
\label{net}       
\end{figure}

Figure~\ref{net} shows preliminary results on fluctuations of net-charge in inelastic p+p interactions~\cite{Mackowiak:2016}. Both the scaled variance $\omega$ and $S\sigma$ depend weakly on collision energy whereas $\kappa\sigma^{2}$ rises significantly. Measured net-charge fluctuations do not agree with independent particle production (Skellam distribution). The difference may come from production of multi-charged particles and quantum statistics~\cite{PBM:2012}. The EPOS 1.99 model provides a reasonably good description of the data.
\section{Summary and conclusion}
NA61/SHINE data taking for the system size $-$ energy scan is well advanced: data for p+p, ${}^{7}$Be+${}^{9}$Be and ${}^{40}$Ar+${}^{45}$Sc collisions have already been recorded. Although preliminary results on transverse momentum and multiplicity fluctuations and higher order moments of the net-charge distribution did not yet give evidence of the critical point of strongly interacting matter, a number of intriguing results were observed, such as suppression of the scaled variance of the multiplicity distribution in nucleus-nucleus collisions, as well as rapid changes in the energy dependence of the kaon inverse slope parameter $T$ and of the ratio $\langle K^{+}\rangle /\langle {\pi}^{+}\rangle$ in p+p interactions. 
\begin{acknowledgement}
This work was supported by the Hungarian Scientific Research Fund
(grants OTKA 68506 and 71989), the J\'anos Bolyai Research Scholarship
of the Hungarian Academy of Sciences, the Polish Ministry of Science
and Higher Education (grants 667\slash N-CERN\slash2010\slash0,
NN\,202\,48\,4339 and NN\,202\,23\,1837), the Polish National Center
for Science (grants~2011\slash03\slash N\slash ST2\slash03691, 
2013\slash11\slash N\slash ST2\slash03879, 
2014\slash13\slash N\slash ST2\slash02565,
2014\slash14\slash E\slash ST2\slash00018
and
2015\slash18\slash M\slash ST2\slash00125), 
the Foundation for Polish Science --- MPD program, co-financed by the
European Union within the European Regional Development Fund, the
Federal Agency of Education of the Ministry of Education and Science
of the Russian Federation (SPbSU research grant 11.38.242.2015), the
Russian Academy of Science and the Russian Foundation for Basic
Research (grants 08-02-00018, 09-02-00664 and 12-02-91503-CERN), the
Ministry of Education, Culture, Sports, Science and Tech\-no\-lo\-gy,
Japan, Grant-in-Aid for Sci\-en\-ti\-fic Research (grants 18071005,
19034011, 19740162, 20740160 and 20039012), the German Research
Foundation (grant GA\,1480/2-2), the EU-funded Marie Curie Outgoing
Fellowship, Grant PIOF-GA-2013-624803, the Bulgarian Nuclear
Regulatory Agency and the Joint Institute for Nuclear Research, Dubna
(bilateral contract No. 4418-1-15\slash 17), Ministry of Education and
Science of the Republic of Serbia (grant OI171002), Swiss
Nationalfonds Foundation (grant 200020\-117913/1), ETH Research
Grant TH-01\,07-3 and the U.S.\ Department of Energy.
\end{acknowledgement}


\begin{thebibliography}{}
%
%
\bibitem{NA61}
N.~Abgrall {\it et al.}, 
JINST {\bf 9}, P06005 (2014)

\bibitem{Fodor:2004nz}
Z.~Fodor and S.D.~Katz,
 JHEP {\bf 0404}, 050 (2004)

\bibitem{Gazdzicki:1998vd}
M.~Gazdzicki and M.I.~Gorenstein,
Acta Phys. Polon. B {\bf 30}, 2705 (1999)

\bibitem{Alt:2007aa}
C.~Alt {\it et al.},
Phys. Rev. C {\bf 77}, 024903 (2008)

\bibitem{Becattini:2006}
F.~Becattini {\it et al.},
Phys. Rev. C {\bf 73}, 044905 (2006)

\bibitem{Lewicki:2016}
M.~Lewicki,
in proceedings of CPOD2016, Wroclaw, Poland (2016)

\bibitem{Pierog:2009zt}
T.~Pierog and K.~Werner,
Nucl.Phys.Proc.Suppl. {\bf 196}, 102-105 (2009)

\bibitem{Abgrall:2013qoa}
N.~Abgrall {\it et al.},
Eur.Phys.J. C {\bf 74}, 2794 (2014)

\bibitem{Naskret:2016}
M.~Naskret,
in proceedings of CPOD2016, Wroclaw, Poland (2016)

\bibitem{Broniowski:2007nz}
W.~Broniowski, M.~Rybczynski and P.~Bozek,
Comput. Phys. Commun. {\bf 180}, 69 (2009) 

\bibitem{Bialas:1976ed}
A.~Bialas, M.~Bleszynski and W.~Czyz,
Nucl.Phys. B {\bf 111}, 461 (1976)

\bibitem{Gazdzicki:2010iv}
M.~Gazdzicki, M.~Gorenstein and P.~Seyboth, 
Acta Phys.Polon. B {\bf 42}, 307-351 (2011)

\bibitem{Golokhvastov:2001}
A.I.~Golokhvastov,
Phys. Atom. Nucl. {\bf 64}, 1841 (2001)

\bibitem{Abbas:2013}
E. Abbas {\it et al.},
Phys. Lett. B {\bf 726}, 610 (2013)

\bibitem{Pulawski:2015tka}
S.~Pulawski,
PoS(CPOD2014)010 (2015) 

\bibitem{Aduszkiewicz:2015dmr}
A.~Aduszkiewicz {\it et al.},
arXiv:1510.03720 [hep-ex] (2015)

\bibitem{Stroebele:2016}
H.~Stroebele
in proceedings of SQM2016, Berkeley, USA (2016)

\bibitem{Gorenstein:2011vq} 
M.I.~Gorenstein, M.~Gazdzicki
Phys. Rev. C {\bf 84} 014904 (2011)

\bibitem{Seryakov:2016}
A.~Seryakov,
in proceedings of CPOD2016, Wroclaw, Poland (2016)

\bibitem{Andronov:2016}
E.~Andronov,
in proceedings of CPOD2016, Wroclaw, Poland (2016), arXiv:1610.05569 [nucl-ex]

\bibitem{Karsch:2011}
F.~Karsch {\it et al.},
Phys. Lett. B {\bf 695}, 136 (2011)

\bibitem{Mackowiak:2016}
M.~Mackowiak-Pawlowska,
in proceedings of CPOD2016, Wroclaw, Poland (2016), arXiv:1610.03838 [nucl-ex]

\bibitem{PBM:2012}
P.~Braun-Munzinger  {\it et al.},
Nucl. Phys. A {\bf 880}, 48 (2012)
\end{thebibliography}
\end{document}